\documentclass{article}

\usepackage{latexsym}

\newcommand{\RR}{\mbox{${\rm \:  R\!\!\!\! I
\;\;}$}}

\newtheorem{theorem}{Theorem}
\newtheorem{itlemma}{Lemma}[section]
\newtheorem{itproposition}[itlemma]{Proposition}
\newtheorem{itcorollary}[itlemma]{Corollary}
\newtheorem{itremark}[itlemma]{Remark}
\newtheorem{itremarks}[itlemma]{Remarks}
\newtheorem{itdefinition}[itlemma]{Definition}
\newtheorem{itexample}[itlemma]{Example}

\newenvironment{lemma}{\begin{itlemma}\rm}{\end{itlemma}} 
\newenvironment{remark}{\begin{itremark}\rm}{\end{itremark}} 
\newenvironment{remarks}{\begin{itremarks} \rm}{\end{itremarks}}
\newenvironment{corollary}{\begin{itcorollary}\rm}{\end{itcorollary}}
\newenvironment{proposition}{\begin{itproposition}\rm}{\end{itproposition}}
\newenvironment{definition}{\begin{itdefinition}\rm}{\end{itdefinition}}
\newenvironment{example}{\begin{itexample}\rm}{\end{itexample}}
\newenvironment{fact}{\noindent {\em Fact}. \ \ }{\hfill \medskip}
\newenvironment{proof}{\noindent {\em Proof}.\ \
}{\hspace*{\fill}$\Box$\medskip}
\newenvironment{claim}{\noindent {\em Claim}. \ \ }{\hfill \medskip}
\newcommand{\be}[1]{\begin{equation}\label{#1}}
\newcommand{\ee}{\end{equation}}
\newcommand{\bl}[1]{\begin{lemma}\label{#1}}
\newcommand{\br}[1]{\begin{remark}\label{#1}}
\newcommand{\brs}[1]{\begin{remarks}\label{#1}}

\newcommand{\bt}[1]{\begin{theorem}\label{#1}}
\newcommand{\bd}[1]{\begin{definition}\label{#1}}
\newcommand{\bp}[1]{\begin{proposition}\label{#1}}
\newcommand{\bc}[1]{\begin{corollary}\label{#1}}
\newcommand{\bfact}[1]{\begin{fact}\label{#1}}
\newcommand{\bex}[1]{\begin{example}\label{#1}}
\newcommand{\ec}{\end{corollary}}
\newcommand{\efact}{\end{fact}}
\newcommand{\eex}{\end{example}}
\newcommand{\el}{\end{lemma}}
\newcommand{\er}{\end{remark}}
\newcommand{\ers}{\end{remarks}}
\newcommand{\et}{\end{theorem}}
\newcommand{\ed}{\end{definition}}
\newcommand{\ep}{\end{proposition}}
\newcommand{\epr}{\end{proof}}
\newcommand{\bpr}{\begin{proof}}
\newcommand{\bcl}{\begin{claim}}
\newcommand{\ecl}{\end{claim}}

\newcommand{\bi}{\begin{itemize}}
\newcommand{\ei}{\end{itemize}}
\newcommand{\ben}{\begin{enumerate}}
\newcommand{\een}{\end{enumerate}}
\newcommand{\text}[1]{\hbox{\rm \ #1\ \/}}

\begin{document}

\begin{center}

{\Large Further results on the observability of quantum systems
under general measurement}

{Domenico D'Alessandro and Raffaele Romano}
\footnote{The authors are with the Department of Mathematics at Iowa
State University, Ames IA-50011
        {\tt\small daless@iastate.edu} {\tt\small rromano@iastate.edu}}

\end{center}




\begin{abstract}

\noindent In this paper,  we present a collection of results on the
observability of quantum mechanical systems,  in the case  the
output is the result of a discrete nonselective measurement. By
defining an {\it effective observable}, we extend previous results,
on the Lie algebraic characterization of observable systems,  to
general measurements.  Further results include the characterization
of a `best probe' (i.e. a minimally disturbing probe) in indirect
measurement and a study of the relation between disturbance and
observability in this case. We also discuss how the observability
properties of a quantum system relate to the problem of state
reconstruction. Extensions of the formalism to the case of selective
measurements are also given.

\end{abstract}

\section{Introduction}

The {\it structural properties} of controllability and
observability have been studied in depth for deterministic control
systems (see e.g. \cite{Sontag}) of the form

\begin{equation} \label{RO1}
\dot x=f(t,x,u),
\end{equation}

\noindent with output

\begin{equation} \label{RO2}
y=y(x).
\end{equation}

In (\ref{RO1}) (\ref{RO2}), $x$ is the state of the system varying
on a given manifold $M$, $u$ is the control, $f$ a smooth vector
field and $y$ a smooth map $M \rightarrow \RR$ which models how
observations on the system depend on the state. For quantum systems,
the study of controllability has received greater attention (see
e.g. \cite{confraIEEE}, \cite{Tarn}, \cite{JS}, \cite{Rama}). A
study of the observability for quantum systems is complicated by the
fact that,  in general, the output has a probabilistic nature and
the associated probability  distribution depends on the current
state. Moreover, different types of measurements can be considered
according to the specific experimental situation at hand. In the
standard text-book selective Von Neumann-Luders  measurement (see
e.g. \cite{Sakurai}),  the measured quantity is represented by a
Hermitian operator $S$ and the result of the measurement is given by
an eigenvalue of $S$ with probability depending on the current
state. However several different scenarios and mathematical models
of quantum measurements can be considered in different situations
(see e.g. \cite{Petruccione}). Therefore different definitions of
observability may be appropriate and of physical interest in
different cases. Nevertheless, there are several reasons to study
observability for quantum mechanical control systems. From the
viewpoint of the fundamental development of the theory,
observability is one of the main concepts to be extended to quantum
systems. It is related to the notion  of input-output equivalence
and therefore to the general question of modeling time varying
Hamiltonians\footnote{Two models are input output equivalent if they
produce the same output function for any input. Two input-output
equivalent models cannot be distinguished  by applying  control
inputs and observing the output and therefore modeling via
input-output experiments may only be made up to equivalence classes
of input-output equivalent models. This question is explored for
networks of particles with spin in \cite{confraCDC},
\cite{confrappmult}}. The problem of determining the state from the
observation of a quorum of observables is an important  one in
quantum mechanics \cite{Pauli}. Techniques to find a set of
observables which would determine the state without ambiguity have
been extensively studied in quantum physics (see e.g. \cite{Amiet},
\cite{infocomp}). Observability of quantum systems is also
particularly important in view of the recent interest in
implementing feedback at the quantum level (see e.g.
\cite{Belavkin}, \cite{MabuchiScience}, \cite{Korotkov},
\cite{Wisemanthesis}, \cite{WisemanPRL}). A feedback controller uses
the knowledge on the current state to update the value of the
control, i.e. it is of the form $u = u(t,x)$. The knowledge of the
state is obtained through the output and therefore an a priori
knowledge of the extent to which information on the state can be
obtained from the output is essential in the design of state
feedback control scheme.

In a recent paper \cite{MikoJPMG}, a study was presented on the
observability properties of quantum systems subject to {\it
nonselective  measurement} i.e. a measurement where either the
result is not read or it is given by the expectation value of a
given observable. The latter case is of interest in several
experimental scenarios such as nuclear magnetic resonance where the
output signal is averaged over a large number of quantum systems. In
these cases,  the definition and treatment of observability is
simplified by the fact that one does not have to consider
probabilities explicitly and natural definitions of observability
can be given. In this paper we expand upon the treatment of
\cite{MikoJPMG} for general measurements. A unified treatment for
the various types of measurements is presented using  notions of
generalized measurement theory \cite{Petruccione}.

We shall be interested in the dynamics of finite dimensional
quantum systems whose state is described by a  density matrix
$\rho$. We shall consider measurements occurring at discrete
instants of time. In between two measurements, the evolution of
$\rho$ is governed by Liouville's equation (see e.g.
\cite{Sakurai})

\begin{equation}\label{RO3}
i \dot{\rho} =[H(u(t)),\rho],
\end{equation}

\noindent where the Hamiltonian $H$ explicitly depends on a control
$u=u(t)$. In general, for nonselective measurement
the result can be assumed to be a linear
function of the current state $\rho$. This is the case when one
performs a Von Neumann-Luders measurement of the expectation value of a
given observable $S$ in which case the output $y$ associated to a
system (\ref{RO3}) is given by

\be{RObis1} y=Tr(S\rho).  \ee

\noindent Another example is the indirect measurement discussed in
detail in Section \ref{indirect}. We shall treat the nonselective case
in greater detail and then present some extensions to the selective case
in Section \ref{SE}.

The effect of nonselective  measurements on
the state $\rho$ of the system can be described in general using the
formalism of {operations}
\cite{Petruccione}, \cite{Kraus}. In particular, if $\cal M$ is a
measurable set of possible outcomes, upon measurement  the state
$\rho$ is modified as

\begin{equation}\label{RO20}
\rho \rightarrow {\cal F} (\rho):= \int_{\cal M} \Phi_m (\rho) \,
dm,
\end{equation}

\noindent or

\begin{equation}\label{RO21}
\rho \rightarrow {\cal F}(\rho):= \sum_{m \in \cal M}
\Phi_m(\rho),
\end{equation}

\noindent according to whether  $\cal M$ is a continuous or discrete
set respectively. The super-operators $\Phi_m$ are called {\it
operations} and,  according to Kraus representation theorem
\cite{Kraus}, can be expressed as

\begin{equation}\label{RO22}
\Phi_m(\rho):=\sum_k \Omega_{mk} \rho \, \Omega_{mk}^*,
\end{equation}

\noindent for a countable set of operators $\Omega_{mk}$.

The plan of the paper is as follows. In Section \ref{OG} we  give
the basic definitions and results concerning the observability of
quantum systems,  under general nonselective measurement. The main
concepts and results given  in  this section were already presented
in \cite{MikoJPMG}, however they are summarized here in a more
compact form by introducing {\it effective} observables. In Section
\ref{indirect} we present some results for the special but important
case of {\it indirect} measurement. These  include an expression for
the effective observable in this case and the derivation of the
optimal measurement in terms of minimal disturbance on the state.
This raises the question of whether there is a conflict between
observability and minimal disturbance. In Section \ref{PE}, by a
simple physical example we show this is not the case: in general we
can have observability with a low disturbance of the system. In
Section \ref{Observers} the design of  quantum state reconstruction
is discussed and related to observability. Section \ref{SE} presents
an extension of the formalism to the case of selective measurement.

\section{Observability under general nonselective measurement}
\label{OG}

If the output $y$ of system (\ref{RO3}) is a linear function of the
current state,  as we assume here, it is always possible to express
$y$ as \be{Seff1} y(t)=Tr \bigl(S_{eff} \rho(t) \bigr),  \ee for
some Hermitian matrix $S_{eff}$, which represents an {\it effective
observable}. Without loss of generality, we can assume that
$S_{eff}$ has zero trace since a trace different from zero would
only introduce a constant shift in the value of the output which
does not play any role in our treatment.
Alternatively, we could quotient all the subspaces (the
observability spaces defined in (\ref{RO17}) below) by
 $span \, \{ i {\bf 1} \}$.

Denote by $\rho_k(t,u, \bar \rho)$ the solution of (\ref{RO3}) with
initial condition $\bar \rho$, control $u$ \, at time $t$ after
$k-1$ measurements, where,  at every measurement, the state is
modified as in (\ref{RO20})-(\ref{RO22}). Then,  two states $\bar
\rho_1$ and $\bar \rho_2$ are called {\it indistinguishable in $k$
steps} (or after $k$ measurements) if, for every control $u$ and
time $t$
\begin{equation}\label{RO7bis}
Tr \bigl( S_{eff} \rho_k (t,u, \bar \rho_1) \bigr) = Tr \bigl(
S_{eff} \rho_k (t,u, \bar \rho_2) \bigr).
\end{equation}
A system is called {\it observable in $k$ steps} if
indistinguishability in $k$ steps of $\bar \rho_1$ and $\bar \rho_2$
implies $\bar \rho_1=\bar \rho_2$. A system is called {\it
observable} if it is observable in $k$ steps for some $k$.

As in the study of controllability (cf. \cite{confraIEEE},
\cite{Rama}, \cite{Tarn}) the {\it dynamical Lie algebra} associated
to the quantum system (\ref{RO3}) plays a prominent role. The
dynamical Lie algebra $\cal L$ is defined as the Lie algebra
generated by $span_{u \in {\cal U}} \{ -i H(u) \}$, where $\cal U$
is the set of possible values for the control $u$. In order to
express the conditions for observability in an arbitrary number of
steps, under general nonselective measurement, we associate to the
super-operator $\cal F$ a {\it dual} super-operator ${\cal F}^*$
acting on observables $S$ and defined from the requirement that, for
every $S$ and $\rho$, $Tr({\cal F}^*(S)\rho)=Tr(S{\cal F}(\rho))$.
Then, we define {\it generalized observability spaces} ${\cal V}_k$,
$k=0,1,...,$ recursively as
\begin{equation}\label{RO17}
\begin{array}{c}
{\cal V}_0 := span \{ iS_{eff} \}, \qquad {\cal V}_1:=
\bigoplus_{j=0}^\infty ad_{\cal L}^j{\cal V}_0, \\ \\
{\cal V}_k: = \bigoplus_{j=0}^\infty ad_{\cal L}^j{\cal F}^*({\cal
V}_{k-1}),
\end{array}
\end{equation}
where $ad_{\cal L}^j {\cal V}$ is defined as spanned by all the
repeated Lie brackets $[R_1, [R_2, \ldots, [R_j, iA] \ldots]],$ and
the Lie bracket is taken $j$ times, $R_1,\ldots,R_j \in {\cal L}$
and $iA \in {\cal V}$.

\noindent With these definitions, the main results of
\cite{MikoJPMG} can be summarized as follows.

\bt{Obseos} System (\ref{RO3}) with output $y$ in (\ref{Seff1}) is
observable in $k$ steps if and only if
\begin{equation}\label{RO11}
{\cal V}_k=su(n).
\end{equation}
More in general,  write $\rho = \rho_1 + \rho_2$
where $\rho_1$ is the component of $\rho$ in $i {\cal
V}_k$\footnote{vector space of Hermitian matrices obtained by
multiplying  by $i$ the skew-Hermitian matrices in ${\cal V}_k$} and
$\rho_2$ is the component along $i{\cal V}_k^\perp$ where ${\cal
V}_k^\perp$ is the orthogonal complement of ${\cal V}_k$ in $u(n)$.
Then, we have the following decomposition of the dynamics
\begin{equation}\label{RO13}
\begin{array}{c}
\dot \rho_1=-i[H(u), \rho_1], \\ \\
\dot \rho_2=-i[H(u), \rho_2],
\end{array}
\end{equation}
and we have
\begin{equation}\label{RO15}
y(t):=Tr \bigl( S_{eff} \rho(t) \bigr) = Tr \bigl( S_{eff} \rho_1(t)
\bigr).
\end{equation}
Initial states are indistinguishable in $k$ steps  if and only if
they differ by an element in $i{\cal V}_k^\perp$.

\et

In several interesting scenarios, the measurement scheme has a `{\it
repetition property}'  which can be defined by imposing that the
operators $\Omega_{mk}$ in (\ref{RO22}) satisfy $\Omega_{mk}
\Omega_{rl}=\delta_{mr} \delta_{kl}\Omega_{mk}$, $\forall m,r \in
{\cal M}$ and $\forall k,l$. In these cases
$\Phi_m(\Phi_m(\rho))=\Phi_m(\rho)$ $\,\forall \rho$,  ${\cal
F}^2={\cal F}$, and ${{\cal F}^*}^2={\cal F}^*$. Physically this
means that a second measurement does not modify the state more than
the first one. In these cases, it is easy to show that
\be{inclusions} {\cal V}_{k-1} \subseteq {\cal V}_k \ee so that
states that are indistinguishable in $k$ steps are also
indistinguishable in $k-1$ steps \footnote{The proof uses an
expression of $S_{eff}$ in terms of effects $F_m$ defined  in
Section \ref{SE}. When the output is an expectation value, then
\be{seff} S_{eff}=\sum_{m \in {\cal M}} m F_m. \ee Moreover using
the expression for the effects
\be{} F_m=\sum_k \Omega_{mk}^* \Omega_{mk},  \ee and the repetition
property, one has ${\cal F}^*(S_{eff})=S_{eff}$ and therefore ${\cal
V}_1={\cal F}^*({\cal V}_0)$. ${\cal V}_0 \subseteq {\cal V}_1$ and
by induction one obtains (\ref{inclusions}).} Moreover, because of
the assumption of finite dimensionality, there exists a $k$ such
that ${\cal V}_k={\cal V}_{\bar k}$ for all $\bar k > k$. An example
is the standard Von Neumann-Luders measurement of the observable
$S$. In this case $S_{eff}=S$. Expressing $S$ as \be{essprex}
S=\sum_j \lambda_j \Pi_j,  \ee where the $\lambda_j$'s are the
eigenvalues of $S$ and $\Pi_j$ are the orthogonal projections onto
the corresponding eigenspaces which play the role of
$\Omega_{mk}$'s. $\cal F$ is given by \be{FcalF} {\cal
F}(\rho):=\sum_j \Pi_j \rho \, \Pi_j.   \ee

In order to use the results of Theorem \ref{Obseos} we need to
find an expression for $\cal F$ and $S_{eff}$ which describe the
particular measurement considered. In the following section we
treat in detail the case of {\it indirect measurement}.

\section{Observability under indirect nonselective measurement}
\label{indirect}

In  indirect measurement, the system evolves as in (\ref{RO3})
until it is in a state $\rho_S$ and it is put in contact with a
{\it probe} system whose initial state we denote by $\rho_P$. The
total system of system and probe at the beginning of the
measurement process is in the state
\begin{equation}\label{RO23}
\rho_{TOT}:=\rho_S \otimes \rho_P.
\end{equation}
During the measurement process, of duration $\tau$, the total system
evolves according to an Hamiltonian
\begin{equation}\label{RO24}
H_{TOT}:= H(u) \otimes {\bf 1} + g(t) A \otimes B+{\bf 1} \otimes
{H_P}.
\end{equation}
The term $H_P$ describes the dynamics of the probe system alone.
The term $g(t)A \otimes B$ gives the interaction between probe and
system, where $g(t)$ is nonzero only during the interval
$[0,\tau]$; $\bf 1$ is the identity operator. It is usually
assumed that,  when the interaction is active, it represents the
dominant term in the Hamiltonian $H_{TOT}$. Therefore we shall
first assume
\begin{equation}\label{RO25}
H_{TOT}:=g(t)A \otimes B.
\end{equation}
At the end of the interval $[0,\tau]$, an observable $S$ is measured
on the probe system, or equivalently an observable ${\bf 1} \otimes
S$ is measured on the total system. In the  following proposition we
calculate an expression for $S_{eff}$ with the Hamiltonian
(\ref{RO25}). \bp{Seffind} With the above definitions and notations,
for indirect measurement \be{RO26} S_{eff} = \sum_{k = 0}^\infty A^k
Tr \Bigl( \bigl( ad_{-iB}^k \rho_P \bigr) S \Bigr) \frac{G^k}{k!},
\ee where \be{RO27} G:=\int_0^\tau g(t)dt. \ee \ep

\bpr The solution of (\ref{RO3}) with initial condition $\rho_{TOT}$
in (\ref{RO23}) and Hamiltonian $H_{TOT}$ in (\ref{RO25}) can be
written at time $\tau$ as \be{RO28} \rho_{TOT}(\tau)=e^{-i G A
\otimes B} \rho_S \otimes \rho_P \, e^{i G A \otimes B}. \ee
Expanding, using the Campbell-Baker-Hausdorff formula, this can be
written as \be{RO29} \rho_{TOT}(\tau)=\sum_{k=0}^\infty ad_{A\otimes
-iB}^k \, \rho_S \otimes \rho_P \frac{G^k}{k!}. \ee Now, it is
easily seen by induction on $k$ that every operator $ad_{A\otimes
-iB}^k \, \rho_S \otimes \rho_P$ can be written in the form
\be{RO30} ad_{A\otimes -iB}^k \, \rho_S \otimes
\rho_P=\sum_{j=1}^{2^k}F_j \otimes L_j, \ee where the $F_j$'s are
all operators  of the form \be{RO31} F_j:=A^{k-l} \rho_S A^l, \ee
for some $l$, $0 \leq l \leq k$, and \be{RO32} \sum_{j=1}^{2^k} L_j
= ad^k_{-iB}\, \rho_P. \ee Using (\ref{RO29}) and (\ref{RO30}) in
the expression of the output $y$, we obtain
\begin{eqnarray}
y = Tr \bigl( {\bf 1} \otimes S \rho_{TOT}(\tau) \bigr) = \nonumber \\
Tr \Bigl( {\bf 1} \otimes S \sum_{k = 0}^\infty ad_{A \otimes -iB}^k
\, \rho_S \otimes \rho_P \frac{G^k}{k!} \Bigr) = \\ \sum_{k =
0}^\infty \frac{G^k}{k!} \sum_{j=1}^{2^k}Tr(F_j \otimes SL_j),
\nonumber
\end{eqnarray}
where $F_j$ and $L_j$ are defined in (\ref{RO30}). Using
(\ref{RO31}), (\ref{RO32}) and elementary properties of the trace we
obtain
\begin{eqnarray}
y = \sum_{k = 0}^\infty \frac{G^k}{k!} Tr ( A^k \rho_S)
\sum_{j = 1}^{2^k} Tr_P(SL_j) = \label{RO34} \\
Tr \Bigl[ \Bigl( \sum_{k = 0}^\infty A^k Tr_P(ad_{-iB}^k \, \rho_P
\, S) \frac{G^k}{k!} \Bigr) \rho_S \Bigr]. \nonumber
\end{eqnarray}
As this has to hold for every $\rho_S$, $S_{eff}$, in $y=Tr(S_{eff}
\rho_S)$, is given by (\ref{RO26}). \epr

\noindent We notice some features of the expression of $S_{eff}$
(\ref{RO26}). \br{Rem1} Assume we retain only the terms up to first
order in $G$. This is  reasonable  if the interaction is very quick
and of small magnitude. Then we have \be{RO35} S_{eff} \approx
Tr_P(\rho_PS){\bf 1} +Tr_P([-iB,\rho_P]S)G A, \ee so that, if
$Tr([-iB,\rho_P]S)\not=0$ there is a one to one correspondence, in
first approximation, between the values of the output and the value
of the observable $A$, and therefore we can say that we are
measuring $A$ indirectly. \er \br{Rem2} In the special case where
$S$ and $B$ are {\it canonically conjugate} observables on the
probe, i.e. \be{RO36} [B,S]=i\gamma {\bf 1}, \ee with $\gamma \in
\RR$, the above correspondence between mean values of $S_{eff}$ and
$A$ is exact. This is the case treated in \cite{Petruccione}. In
order to see this, consider the expression of $S_{eff}$
(\ref{RO26}). From the property \be{RO37} Tr_P \bigl(
(ad^k_{-iB}\,\rho_P)S \bigr)=(-1)^kTr \bigl( (ad^k_{-iB}S)\rho_P
\bigr) \ee and (\ref{RO36}), we obtain that all the terms in the sum
corresponding to $k \geq 2$ are zero as $ad^k_{-iB}S$ is zero in
these cases. Therefore $S_{eff}$ reduces to \be{RO38} S_{eff} =
Tr_P(\rho_PS){\bf 1} +\gamma G A, \ee where $\gamma$ is the one in
(\ref{RO36}) and we have used the fact that $Tr(\rho_P)=1$. \er

\br{Rem3} In some cases,  it is not appropriate to neglect the term
containing $H(u)$ in (\ref{RO24}). In these cases, it is not
possible,  in general,  to obtain a simple expression of $S_{eff}$
as in (\ref{RO26}). However Remark \ref{Rem1} above still holds
true, assuming the $g(t)$ is a simple square function in $[0,\tau]$
so that $G = \tau$, and $u$ is constant in $[0,\tau]$. Notice that
we can write,  generalizing (\ref{RO29}), \be{RO39}
\rho_{TOT}(\tau)= \sum_{k=0}^\infty ad_{-iH(u)\otimes {\bf 1}+
A\otimes -iB}^k \, \rho_S \otimes \rho_P \frac{\tau^k}{k!}. \ee and
expanding $\rho_{TOT}(\tau)$ to first order in $\tau$, we have
\be{RO40} \rho_{TOT}\approx \rho_S \otimes \rho_P +
ad_{-iH(u)\otimes {\bf 1}+ A \otimes -iB} \,\rho_S \otimes \rho_P \,
\tau. \ee Calculating $Tr({\bf 1} \otimes S \rho_{TOT})$, using the
fact that $[-iH(u) \otimes {\bf 1}, \rho_S \otimes \rho_P]=[-iH(u),
\rho_S] \otimes \rho_P$, and that $Tr([-iH(u), \rho_S])=0$, we
obtain the same expression for $S_{eff}$ as in (\ref{RO35}),  with
$G$ replaced by $\tau$. \er

\br{Rem4} The expression of $S_{eff}$ does not depend on the probe
being finite dimensional. \er

\noindent The action on $\rho_S$ ($\cal F$ in (\ref{RO20}),
(\ref{RO21})) after an indirect measurement is given by \be{RO41}
\rho_S \rightarrow {\cal F} (\rho_S):= Tr_P (e^{A \otimes {-iB}G}
\rho_S \otimes \rho_P\, e^{A \otimes {iB} G}), \ee and it is
independent of the observable $S$ measured. This is easily seen by
using the properties of the partial trace $Tr_P$ and the expression
of $S$ (\ref{essprex}) in terms of projections. We refer to
\cite{Petruccione} (Section 2.4.6) for a discussion of how this
operation on the state can be rewritten according to Kraus
representation theorem as in (\ref{RO20})-(\ref{RO22}).

In (\ref{RO26}),  there is a  dependence of $S_{eff}$ on the
initial state of the probe. As a consequence,  it could be
possible to modify the observability property for the system by
suitably choosing $\rho_P$. However,  the disturbance induced on
the system depends on $\rho_P$ as well, and it is interesting to
investigate whether  there is a conflict between observability and
low disturbance of the system.

With this motivation in mind, we provide here an analysis of the
disturbance on the state while performing an indirect non selective
measurement and show how to find the initial state of the probe
which gives the (worst case) minimal disturbance. Using this result,
we shall show in the next section, with an example, that there is in
general no conflict between observability and minimal disturbance.

We consider,  as a measure of the disturbance on the state
$\rho_S$,  the trace norm
\begin{equation}
\label{DO1} d  := \| {\cal F}(\rho_S)  - \rho_S (0) \| = \bigl[ Tr
\bigl( {\cal F}(\rho_S )  - \rho_S  \bigr)^2 \,
\bigr]^{\frac{1}{2}},
\end{equation}
expressing the distance between the initial state $\rho_S $ and the
final one, ${\cal F}(\rho_S)$. If we fix all the parameters of the
measurement process, the disturbance $d$ will in general be a convex
function of $\rho_S$. Since $\rho_S$ varies on a convex and compact
set, the set of all the density matrices, the maximum will in
general be achieved on the boundary i.e. it will be a pure state. We
shall now show how  it is possible to find this worst case pure
state in the small time approximation in the case where all the
terms in (\ref{RO24}) are possibly different from zero (and $u$ is
constant). After that, we will derive the corresponding distance
$d$, depending on $\rho_P$. Then, it will be immediate to find the
initial state of the probe which gives the minimum for $d$. In the
above situation, neglecting higher order terms in $\tau$, $d^2$ can
be written as
\begin{equation}
\label{DO3} d^2 = -\tau^2 \, Tr \bigl( [ H(u) + Tr_P (B \rho_P) A,
\rho_S (0) ] \bigr)^2.
\end{equation}
If we set \be{RO42} X := H(u) + Tr_P (B \rho_P) A,  \ee we can write
$d^2$ as \be{RO43} d^2 = 2 \tau^2 Tr ( X^2 \rho_S^2 - X \rho_S X
\rho_S ), \ee where we write $\rho_S$ for $\rho_S(0)$ as there is no
possibility of confusion. As an orthonormal basis for the Hilbert
space of the system,  we choose the eigenvectors of the Hermitian
operator $X$,  $\vert \phi_k \rangle$, $k = 1, \ldots , n$, so
\be{RO00_new} X = \sum_{k = 1}^{n} x_k \vert \phi_k \rangle \langle
\phi_k \vert \ee and $x_k$ are the real eigenvalues of $X$. Since
the worst case $\rho_S$ is a pure state, we can write $\rho_S =
\vert \psi \rangle \langle \psi \vert$ for some $\vert \psi \rangle
= \sum_k r_k \vert \phi_k \rangle$ where the $n$ coefficients $r_k$
completely specify $\rho_S$. They can be assumed real by suitably
redefining the eigenvectors $\vert \phi_k \rangle$. We have the
further constraint $\sum_k r_k^2 = 1$ since $Tr \rho_S = 1$. To
determine the worst case $\rho_S$, we rewrite (\ref{RO43}) as a
function of the $r_k$ coefficients \be{RO01_new} d^2 = 2 \tau^2
\Bigl( \sum_{k = 1}^n x_k^2 r_k^2 - \sum_{kl = 1}^n x_k x_l (r_k
r_l)^2 \Bigr) \ee and, rearranging the terms, \be{RO02_new} d^2 = 2
\tau^2 \Bigl( \sum_{k > j} (x_k - x_j )^2 (r_k r_j)^2 \Bigr). \ee We
now maximize $d^2$ with respect to the $n$ parameters $r_k$ using
the Lagrange method: \be{RO03_new} \left\{
\begin{array}{ll}
    \partial_{r_l} \tilde{f} (r_1, \ldots, r_n) = 0 &
    \hbox{for $l = 1, \ldots , n$} \\ \\
    \partial_{\lambda} \tilde{f} (r_1, \ldots, r_n) = 0, & \hbox{} \\
\end{array}
\right.  \ee where \be{RO04_new} \tilde{f} (r_1, \ldots, r_n) = d^2
+ \lambda \Bigl( \sum_{k = 1}^n r_k - 1 \Bigr) \ee and $\lambda$ is
the Lagrange multiplier. More explicitly, \be{RO05_new} \left\{
\begin{array}{ll}
    r_l \left( \sum_{j \ne l} (x_l - x_j)^2 r_j^2 + \lambda \right) = 0
    & \hbox{for $l = 1, \ldots , n$} \\ \\
    \sum_k r_k^2 = 1. & \hbox{} \\
\end{array}
\right. \ee System (\ref{RO05_new}) always admits a solution since
the function $d^2$ is continuous over the compact set of pure
density matrices. In the next section we will explicitly compute
$\rho_S$ in a particular case. Without solving (\ref{RO05_new}) in
the general case, we summarize our discussion in the following
theorem.

\bt{maxmin} The worst case disturbance in a small time approximation
is given by $d^2$ in (\ref{RO02_new}), where $(r_1, \ldots, r_n)$
are the solution of system (\ref{RO05_new}). Therefore given $u$,
$A$ and $B$ in the definition of $X$, the initial state of the probe
which minimizes the worst case error has to be chosen so as to
minimize this $d^2$. \et

\section{Observability and minimal disturbance}
\label{PE}

As a concrete example of observability under an indirect
measurement, we consider the simple case of two-dimensional system
and probe. The system is a qubit with external control $u$ affecting
a two-components magnetic field, for example \be{PE0} H(u) = E_x (u)
\sigma_x + E_y (u) \sigma_y. \ee We assume a piecewise constant
control $u \in \{u_1, u_2\}$ that flips the magnetic field
directions $x$ and $y$, that is $E_x(u_1) = E$, $E_y(u_1) = 0$ and
$E_x(u_2) = 0$, $E_y(u_2) = E$. We use a second qubit as probe and
we let it interact with the system for a short time $\tau$ in which
the free evolution (\ref{PE0}) can be neglected. To get information
about the initial state $\rho_S$ we measure $S = \sigma_z$ on the
probe. Assuming a simple Ising model of interaction, $A = \sigma_y$
and $B = \sigma_x$, the effective observable $S_{eff}$ can be
explicitly computed. Splitting the sum in (\ref{RO26}) in even and
odd indices, using (\ref{RO37}) and considering that
\be{PE1} A^k = \left\{%
\begin{array}{ll}
    {\bf 1} & \hbox{for $k$ even,} \\ \\
    \sigma_y & \hbox{for $k$ odd} \\
\end{array}%
\right.     \ee \noindent and \be{PE2} ad^k_{-iB} S =
\left\{%
\begin{array}{ll}
    (-1)^{k/2} \, 2^k \, \sigma_z & \hbox{for $k$ even,} \\ \\
    (-1)^{1 + k/2} \, 2^{k + 1} \, \sigma_y & \hbox{for $k$ odd} \\
\end{array}%
\right. \ee we find that \be{PE3} S_{eff} = Tr_P (\sigma_z \rho_P)
\, \cos{2G} \, {\bf 1} + Tr_P (\sigma_y \rho_P) \, \sin(2G) \,
\sigma_y. \ee

\br{Rem5} The observability properties of our system strongly
depend on the initial state of the probe $\rho_P$. Suppose that
$Tr (\sigma_y \rho_P) = 0$; in such a case $S_{eff} = 0$ and the
observability spaces ${\cal V}_k$ contain only the null vector.
Then, for any $k$ the system is not observable and the states are
all indistinguishable. On the other hand,  suppose $Tr (\sigma_y
\rho_P) \ne 0$. In such a case $S_{eff} = Tr (\sigma_y \rho_P) \,
\sin{2G} \, \sigma_y$ and ${\cal V}_k = su(2)$ for all $k$, and
the system is observable in $k$ steps, for every $k$. \er

In some cases,  it is not appropriate to neglect the free evolution
of the system. However, following Remark \ref{Rem1} we can
explicitly evaluate the effective observable: \be{PE4} S_{eff} =
Tr_P (\sigma_z \rho_P) \, {\bf 1} + 2 \tau Tr_P (\sigma_y \rho_P) \,
\sigma_y \ee where $\tau$ is the time of interaction (assumed small)
and $g(t)$ is a square function. Remark \ref{Rem5} holds true in
that case as well.

We now determine the minimal disturbing probe described in Theorem
\ref{maxmin}. We assumed that during the time interval $\tau$ the
control does not change, and its actual value is relevant in order
to find the minimal disturbing probe. In our example,
(\ref{RO05_new}) becomes \be{PE5} \left\{
\begin{array}{ll}
    r_1 \left((x_2 - x_1)^2 r_2^2 + \lambda \right) = 0 & \hbox{} \\
    r_2 \left((x_2 - x_1)^2 r_1^2 + \lambda \right) = 0 & \hbox{} \\
    r_1^2 + r_2^2 = 1. & \hbox{} \\
\end{array}
\right.     \ee where $x_1$, $x_2$ are the eigenvectors of $X$ and
they depend on $u$. Solving (\ref{PE5}) we find the worst case
$\rho_S$:
\begin{equation}\label{PE6}
    \rho_S = \frac{1}{2} \left( \vert \phi_1 \rangle \langle \phi_1 \vert +
    \vert \phi_2 \rangle \langle \phi_2 \vert \pm
    \vert \phi_1 \rangle \langle \phi_2 \vert \pm
    \vert \phi_2 \rangle \langle \phi_1 \vert \right)
\end{equation}
leading to $d^2 = (x_2 - x_1)^2 /4$. For $u = u_1$,  $x_2 - x_1 =
2[E^2 + (Tr_P (\sigma_x \rho_P))^2]$, for $u = u_2$, $x_2 - x_1 =
2(E + Tr_P (\sigma_x \rho_P))$. Then, the minimally disturbing
probe must satisfy \be{PE7} \left\{
  \begin{array}{ll}
    Tr_P (\sigma_x \rho_P) = 0 & \hbox{for $u = u_1$,} \\ \\
    Tr_P (\sigma_x \rho_P) =\max \{-E, -1 \}& \hbox{for $u = u_2$.}
  \end{array}
\right. \ee In both cases there is not a conflict between
observability and minimal disturbance (see Remark \ref{Rem5}).

\section{Observability and state reconstruction}
\label{Observers}

We present in this section a system theoretic treatment of the
problem of state determination for the system (\ref{RO3}) with
output (\ref{RObis1}). In systems and control theory, for a
continuous time system such as (\ref{RO3}),  under observability
conditions, the (initial) state is determined from a continuous
reading of the output.  From a physics point of view, a continuous
monitoring of the output will introduce a back action on the state
of the quantum system and therefore it will render invalid the model
(\ref{RO3}). However, this scheme is  of interest for quantum
systems in situations like the following. Assume we want to
determine the unknown (initial) state and we have many copies of the
same  system. We perform a nonselective measurement on each copy at
slightly different times so as to simulate a continuous measurement.
The data so obtained can then be used by the observer to reconstruct
the state of the system (without measurement back-action).

With this motivation in mind, a method for reconstructing the
initial state can be obtained by adapting to our case techniques for
time varying linear systems \cite{Kailath}. Observability (in one
step) is a necessary and sufficient condition for reconstructing the
initial state from a reading of the output. In fact, if the system
is not observable, then it is not possible to discern between two
indistinguishable initial states. Viceversa, assume the system is
observable. Then, we have that \cite{MikoJPMG} \be{equi} \{X^* iS X
| X \in e^{\cal L}\}=su(n). \ee This means that we can choose a
control $u$, so that, for the corresponding solution $X_u$ of
Schr\"odinger operator equation \be{Scrop} \dot X=-iH(u)X, \quad
 X(0)=I,
 \ee
 the $n^2-1$ elements of the matrix
$X_u^*SX_u$ (namely the real functions composing the matrix modulo
the fact that this matrix is Hermitian) are linearly independent.
$e^{\cal L}$ is the Lie group  of all the matrices $X_m$ for which
there exists a control steering $X$ in (\ref{Scrop}) from the
identity to $X_m$. We  can select $n^2-1$ matrices
$X_1,...,X_{n^2-1}$ so that $X^*_1 S
X_1$,...,$X_{n^2-1}^*SX_{n^2-1}$ are linearly independent and then
concatenate the controls steering the matrix $X$ in (\ref{Scrop}) to
$X_1$, $X_2X_1^*$, $X_3 X_2^*$,...,$X_{n^2-1}X_{n^2-2}^*$.
Now assume that, in the control interval $[0,T]$, the (significant)
real entries of $X_u^*SX_u$ are linearly independent and define the
linear operator $\cal W$ which maps $n \times n$ Hermitian matrices
with zero trace  into $n \times n$ Hermitian matrices with zero
trace as follows \be{Wrosk} {\cal W}_u(\hat{\rho_0}):=\int_0^T
X_u^*(t)S X_u(t)Tr \bigl(X^*_u(t)SX_u(t) \hat{\rho_0} \bigr)dt. \ee
The operator ${\cal W}_u$ has the following property. \bp{propW} If
the $n^2-1$ real functions composing $X_u^*SX_u$ are linearly
independent then ${\cal W}_u$ has rank $n^2-1$ and therefore it has
an inverse ${\cal W}_u^{-1}$ \ep \bpr This follow from the well
known fact that (see e.g. \cite{Kailath} Section 9.2.1) $m$
functions $l_j=l_j(t)$, $j=1,...,m$ are linearly independent in an
interval $[0,T]$ if and only if the matrix \be{Gram}
g_{ij}:=\int_0^T l_i(t) l_j(t) dt \ee is nonsingular. In our case,
if we order the $n^2-1$ elements of $\hat{\rho_0}$ by row and then
with real and imaginary part i.e. as $\hat{\rho_0}_{1,1}$, $Re(\hat
{\rho_0}_{1,2})$, $Im(\hat {\rho_0}_{1,2})$,...,$Re(\hat
{\rho_0}_{(n-1),n})$, $Im(\hat {\rho_0}_{(n-1),n}),\hat
{\rho_0}_{n,n}$ and the entries of $X_u^*SX_u$ in the same way, the
matrix which represents the linear application ${\cal W}_u$ has the
form (\ref{Gram}) where $l_i$ are the elements of $X_u^*SX_u$ and
therefore it is invertible. \epr

\noindent Now, from formula (\ref{RObis1}), we obtain \be{stand1}
y(t)=Tr \bigl(X^*_uSX_u(\rho_0- \frac{1}{n}I_{n \times n}) \bigr),
\ee and therefore \be{stand2} \int_0^TX^*_u(t) S X_u(t)  y(t) dt =
\int_0^T X^*_u(t) S X_u(t)Tr\bigl(X^*_uSX_u(\rho_0- \frac{1}{n}I_{n
\times n}) \bigr) dt. \ee Therefore, using the definition of ${\cal
W}_u$ (\ref{Wrosk}), we have the following formula for the
reconstruction of the initial state $\rho_0$, \be{reconst} \rho_0
=\frac{1}{n} I_{n \times n}+ {\cal W}_u^{-1} \left( \int_0^T
X^*_u(t) S X_u(t) dt \right).  \ee Formula (\ref{reconst})
represents a system theoretic alternative to methods for quantum
state tomography. We summarize the discussion in the following
theorem.

\bt{Finale} Consider  system (\ref{RO3}) with output (\ref{RObis1}).
If the system is observable (in one step), then there exists a
control such that formula (\ref{reconst}) gives the initial state.
\et

An alternative to the '{\it static}' state reconstruction formula
(\ref{reconst}) is the design of an {\it asymptotic observer} namely
a dynamical system which uses only a reading of the output and whose
state asymptotically converges to the actual state of the system. We
present  in the rest of this section proposal for such an asymptotic
observer which is inspired the treatment for linear time varying
systems in \cite{Jap}. We notice that for static state
reconstruction we imposed a requirement on the control which impled
that the operator defined in formula (\ref{Wrosk}) has full rank
(cf. Proposition \ref{propW}). For an asymptotic observer which
estimates the state as $t \rightarrow \infty$, we need to impose
that this property is somehow uniform for every $t$ as $t
\rightarrow \infty$. To make this more precise we define a time
dependent, linear symmetric operator on Hermitian matrices ${\cal
P}_t$ as follows (we omit for notational simplicity the dependence
on the control $u$). Let $U:=U(t)$  be the solution of Schr\"odinger
operator equation \be{Scrop6} \dot U=iH(u)U, \qquad U(0)=I, \ee
where $H(u)$ is the same as in (\ref{RO3}) and (\ref{Scrop}). We
define \be{PT} {\cal P}_t(\Delta):={\cal P}_t^{M,\sigma}(\Delta):=
\int_{t - \sigma}^t e^{-M(t-\tau)} U(t) U^*(\tau)SU(\tau) U^*(t)
Tr\bigl( U(t) U^*(\tau)SU(\tau) U^*(t) \Delta \bigr) d \tau, \ee
with $\sigma
>0$ and $M
>0$. We assume that the control $u$ is such that there exists a
$\sigma >0$, such that, for every $t \geq \sigma$ \be{con67}
\alpha_1 Tr(\Delta^2) \leq \int_{t - \sigma}^t \left( Tr \bigl( U(t)
U^*(\tau)SU(\tau) U^*(t) \Delta \bigr) \right)^2d \tau  \leq
\alpha_2 Tr(\Delta^2), \ee for some positive constants $\alpha_1$
and $\alpha_2$.

\noindent We choose the same $\sigma$ in the definition (\ref{PT})
and our assumption implies that \be{Ineq}
 \alpha_1 e^{-M\sigma} Tr(\Delta^2) \leq Tr(\Delta {\cal P}_t(\Delta)) \leq
 \alpha_2 Tr(\Delta^2),
\ee and that ${\cal P}_t$ is nonsingular, so that we can define the
inverse operator ${\cal P}_t^{-1}$. Moreover from the definition
(\ref{PT}), we obtain ($\Delta$ constant) \be{derivPT} \frac{d}{dt}
{\cal P}_t(\Delta)= STr(S\Delta) -e^{-M\sigma} U(t)U^*(t-\sigma)S
U(t-\sigma)U^*(t) Tr\bigl(U(t)U^*(t-\sigma)S U(t-\sigma)U^*(t)\Delta
\bigr) \ee
$$-M{\cal P}_t(\Delta) +[iH(u), {\cal P}_t (\Delta)] +{\cal P}_t([iH(u), \Delta]). $$


We consider the state observer for system (\ref{RO3}) \be{Statob1}
\frac{d}{dt} \hat \rho= [-iH(u), \hat \rho] - \frac{1}{2}{\cal
P}_t^{-1}(S)Tr\bigl(S(\rho - \hat \rho)\bigr)= [-iH(u), \hat \rho] -
\frac{1}{2}{\cal P}_t^{-1}(S)(y-Tr(S\hat \rho)). \ee Here $\hat
\rho$ is the estimate of the actual state $\rho$ and we are going to
show that $\Delta(t):=\rho(t)-\hat \rho(t)$ tends to zero as $t
\rightarrow \infty$. By subtracting (\ref{Statob1}) from
(\ref{RO3}), we obtain the differential equation for $\Delta$,
\be{DiffDelta} \dot \Delta= [-i H(u), \Delta]-\frac{1}{2}{\cal
P}_t(S)Tr(S \Delta). \ee

 Define the Lyapunov candidate function $V:=V(t,\Delta)=Tr(\Delta
 {\cal P}_t(\Delta))$, which according to (\ref{Ineq}) satisfies
 \be{Liapo1}
\alpha_1' Tr(\Delta^2) \leq V(t,\Delta) \leq \alpha_2' Tr(\Delta^2),
 \ee
for appropriate positive constants $\alpha_1'$ and $\alpha_2'$.
Moreover we can calculate $\frac{d}{dt} V(t, \Delta(t))$. By using
(\ref{DiffDelta}) and (\ref{derivPT}) along with the property
$Tr({\cal P}_t(\Delta) {\cal P}_t^{-1}(S))=Tr(\Delta S)$,
we obtain \be{finalV} \frac{d}{dt}V(t,\Delta) \leq -M V(t,\Delta),
\ee and therefore it follows from Lyapunov second method \cite{Hahn}
that system (\ref{DiffDelta}) is asymptotically stable, and
therefore $\Delta$ tends to zero as $t \rightarrow \infty$. Notice
that our Lyapunov is only defined for $t$ sufficiently large ($t
\geq \sigma$) however this does not change the stability analysis as
all the quantities considered are guaranteed to be bounded over a
finite interval of time. We conclude with the following Theorem.
\bt{Finale8} Consider system (\ref{RO3}) with output (\ref{RObis1}).
Assume that the control $u$ satisfies the condition (\ref{con67}).
Then system (\ref{Statob1}) is an asymptotic observer for
(\ref{RO3}). \et

\section{Some extensions to selective measurement}
\label{SE}

In this section,  we discuss how the theory described above for
nonselective measurement extends to selective measurement. There is
no difficulty in doing this in the most general case namely in the
context of the generalized measurement theory of operations and
effects \cite{Petruccione}. According to this theory, given a
measurement scheme, to every result  $m$ is associated a positive
operator $F_m$, called an {\it effect}. If $\rho$ is the current
state of the system, the probability of obtaining the result $m$ (or
of an event $m$ to occur) is \be{obtainm} P(m)=Tr(F_m \rho). \ee
After a result $m$ (or, more generally an event $m$) has occurred,
the state is modified according to \be{statemod} \rho \rightarrow
P(m)^{-1}\Phi_m(\rho), \ee where the positive super-operators
$\Phi_m$ are the same  {\it operations} as in (\ref{RO22}) and
$Tr(\Phi_m (\rho))=P(m)=Tr(F_m \rho)$. Two initial states $\bar
\rho_1$ and $\bar \rho_2$ are said to be indistinguishable in $k$
steps, in selective measurement,  if they give every possible result
with the same probability at the $k-$th measurement, for every
choice of the control $u$. In formulas (cf. (\ref{RO7bis}))
\begin{equation}\label{RO7tris} Tr(F_m\rho_k(t,u,
\bar \rho_1))=Tr\bigl(F_m\rho_k(t,u, \bar \rho_2)\bigr) \qquad
\forall m \in {\cal M},
\end{equation}
where $\cal M$ is the set of possible results (events). Let $P_k(m)$
be the probability of having the result $m$ at the $k-$th
measurement and let $P(m_1,...,m_k)$ be the joint probability of
having result $m_1$ at the first step, $m_2$ at the second step and
so on. Also, indicate by $P_k(m_k|m_1,\ldots m_{k-1})$ the
conditional probability of having $m_k$ at the $k-$th measurement,
given $m_1$,\ldots $m_k$ as ordered results of the previous
measurements. By use of the formula \be{prob1}
P_k(m)=\sum_{m_1...m_{k-1}}P_k(m|m_1,..., m_{k-1}) P(m_1,...,
m_{k-1}), \ee and repeated use of Bayes' formula
\be{prob2}
P(m_1,...,m_{k-1})=
\ee
$$P_{k-1}(m_{k-1}|m_1,...,m_{k-2})P(m_1,...,m_{k-2}),$$
we can write $P_k(m)$ starting from an initial condition $\rho_0$ as
\begin{eqnarray}
\label{prob4} P_k(m)= \sum_{m_1...m_{k-1}} Tr \Bigl( F_m X_k
(\Phi_{m_{k-1}} (X_{k-1} (\Phi_{m_{k-2}} \nonumber \\
\ldots ( \Phi_{m_1}(X_1 \rho_0 X_1^*))\ldots ) ) X_{k-1}^* )) X_k^*
\Bigr)
\end{eqnarray} where $X_j$, $j=1,\ldots,k$ is  the evolution solution
of the Schr{\"o}dinger operator equation (\ref{Scrop}) in the
interval between the $(j-1)$-th measurement and the $j-$th
measurement. Using (\ref{prob4}) and using the linearity of the
operators $\Phi_m$, we can rewrite $P_k(m)$ as \be{prob6} P_k(m)=Tr
\Bigl( F_m X_k {\cal F}( X_{k-1}... {\cal F}(X_1 \rho_0 X_1^*)...
X_{k-1}^*) X_k^* \Bigr) \ee where $\cal F$ is defined in
(\ref{RO21}). From this point on the theory goes as in
\cite{MikoJPMG} and the result is an extension of Theorem
\ref{Obseos}. In particular,  one defines the `selective'
observability spaces (cf. (\ref{RO17}))
\begin{equation}\label{RO17sel}
\begin{array}{c}
{\cal V}_0^{sel}:= span_{m \in {\cal M} }\{ iF_m \}, \qquad {\cal
V}_1^{sel}:=\bigoplus_{j=0}^\infty ad_{\cal L}^j({\cal
V}^{sel}_{0}), \\ \\
{\cal V}_k^{sel}:=\bigoplus_{j=0}^\infty ad_{\cal L}^j{\cal
F}^*({\cal V}^{sel}_{k-1}),
\end{array}
\end{equation}
and Theorem \ref{Obseos} extends by replacing nonselective
observability with selective observability and the spaces $\cal V$
with the spaces ${\cal V}^{sel}$
\footnote{Condition (\ref{}) of Theorem \ref{Obseos} needs to be
slightly modified as the effects $F_m$ do not necessarily have zero
trace, by replacing ${\cal V}_k$ with ${\cal V}_k/{ span  \{ i {\bf
1} \} }$ or by making all the effects traceless.}.


The remarks following Theorem \ref{Obseos} on the implications of
the repetition property also extend with only minor formal
modifications. In the particular case of the standard Von
Neumann-Luders measurement,  the observable $S$ is written in terms
of the projectors $\Pi_\lambda$ and the eigenvalues $\lambda$ as
\be{hhjh} S=\sum_{\lambda \in {\cal M}} \lambda \Pi_\lambda,  \ee
and the above theory holds with $\Pi_\lambda$ playing the role of
the effects $F_m$.

\br{SobsVSNonSobs} The observability space ${\cal V}_0^{sel}$ does,
in general, include the observability space  ${\cal V}_0$ and
therefore the same is true for the observability spaces ${\cal
V}_k^{sel}$ and ${\cal V}_k$. This implies that nonselective
observability implies selective observability, as it is intuitive
but not viceversa. Consider as a specific example a spin $1/2$
particle for which the $z$ component of the spin is measured, with a
Von Neumann-L\"uders measurement. In this case ${\cal V}_0=span \,\{
i \sigma_z \}$, while ${\cal V}_0^{sel}= span \,  \{  i \sigma_z, i
{\bf 1} \}$, with $\sigma_z$ the $z$-Pauli matrix. So up to the span
of $i {\bf 1}$ the two subspaces are the same and the observability
properties in the selective and non selective case are the same and
depend on the dynamics. However consider a spin $1$ on  which we
perform a Von Neumann-L\"uders measurement of the spin along the $z$
direction. In this case, using the representations of the spin
angular momentum calculated for a spin $1$ (see e.g. \cite{Sakurai}
Section 3.5), we have \be{V0NS} {\cal V}_0=span \, \{ \pmatrix{i & 0
& 0 \cr 0 & 0 & 0 \cr 0 & 0 & -i} \}, \ee and \be{V0S} {\cal
V}_0^{sel}/{span \{ i{\bf 1} \}} = span \,\{ \pmatrix{i & 0 & 0 \cr
0 & -i & 0 \cr 0 & 0 & 0}, \quad \pmatrix{0 & 0 & 0 \cr 0 & i & 0
\cr 0 & 0 & -i} \}.  \ee Therefore ${\cal V}_0$ and ${\cal
V}_0^{sel}/ span { \{i {\bf 1} \}} $ do not coincide in this case.
In particular, if we consider the dynamics determined by a
time-varying control electro-magnetic field in the $x-y$ plane, we
have that ${\cal V}_1$ is spanned by the three dimensional Lie
algebra representation of $su(2)$, while ${\cal V}_0^{sel}/ {span
\{i {\bf 1} \}} $ is equal to $su(3)$, as one can easily verify by
repeated Lie brackets. Therefore, in this case, we have selective
observability but not non selective observability, in one step. The
situation is the same if we consider observability in $k$ steps as
${\cal V}_1={\cal V}_k$ for every $k$.
 \er

\section{Conclusions}
This paper has presented a collection of results on the
observability of quantum systems with emphasis on the case of
nonselective measurement. In particular

\begin{enumerate}

\item Using the formalism of generalized measurement and of effects
and operations we have extended the basic definitions and criteria
of observability to the case of general measurement by introducing
an effective observable.

\item We have derived a general  expression for the effective
observable in the case of Von Neumann indirect measurement.

\item In the case of indirect measurement, we have derived an
expression for the state of the probe which would introduce the
minimum disturbance in the state to be measured. We have showed that
the requirement of a minimal disturbing probe does not in general
compromise the observability properties of the resulting system and
therefore the amount of information obtained on the state by the
measurement of the output.

\item We have presented two system theoretic methods to reconstruct
the state by a measurement of the expectation value of an
appropriate observable.   One of them is through an integral formula
and uses readings over a finite interval of time. The other is
through an asymptotic observer whose state converges to the state of
the measured system.

\item We have extended the basic definitions and observability
criteria to selective measurements.

\end{enumerate}

We believe that the system theoretic approach to quantum  state
determination is worth being further investigated. Extensions of our
definitions and results to continuous measurements, optimization of
the methods for state determination in specific settings,
applications of observer design in closed loop quantum systems are
only few possible subjects for future research.

{\bf Acknowledgment} {This work was supported by NSF under Career
grant ECS-0237925}

\end{document}